\documentstyle[12pt,prl,aps,epsfig,citesort,psfig19,floats]{revtex}
\renewcommand {\baselinestretch}{1.}

\newcommand{\aleq}{\mbox{\ 
\raisebox{-.9ex}{$\stackrel{\textstyle<}{\sim}$}\ }}
\newcommand{\ageq}{\mbox{\
\raisebox{-.9ex}{$\stackrel{\textstyle >}{\sim}$}\ }}

\def\x{{\mbox{\boldmath$x$}}}
\def\u{{\mbox{\boldmath$u$}}}
\def\r{{\mbox{\boldmath$r$}}}

\def\e{{\mbox{\boldmath$e$}}}

\def\dz{{\delta\zeta}}
\def\dx{{\delta\xi}}
\def\la{{\langle}}
\def\ra{{\rangle}}

\def\eps{{\epsilon}}

\def\rel{{Re_\lambda}}

\def\begineq{\begin{equation}}
\def\endeq{\end{equation}}
\def\be{\begin{equation}}
\def\ee{\end{equation}}

\begin{document}
\bibliographystyle{prsty}
%\psdraft

\title{
Application of  extended
self similarity in turbulence
}
\author{Siegfried Grossmann,
Detlef Lohse, and
Achim Reeh}
\address{
Fachbereich Physik der Universit\"at Marburg,\\
Renthof 6, D-35032 Marburg, Germany}

\date{\today}

\maketitle
\begin{abstract}
From Navier-Stokes turbulence numerical simulations we show that
for the extended self similarity (ESS) method it 
is essential to take the third order structure function
{\it taken with the modulus} and called $D_3^*(r)$,
rather than the standard third order
structure function $D_3(r)$ itself.
If done so, we find ESS towards scales larger than order $ \sim 10\eta$,
where $\eta$ is the Kolmogorov scale.
If $D_3(r)$ is used,
there is no ESS. We also analyze ESS within 
the Batchelor parametrization of the second
and third order longitudinal
structure function and 
focus on the scaling of the transversal structure function.
The Re-asymptotic inertial range
scaling develops  only  beyond a Taylor-Reynolds number
$\rel \ageq 500$.
\end{abstract}

%----------------------------------------------------------------------

\section{Introduction}
Extended self similarity (ESS, \cite{ben93b,ben94a,bri94})
has been most useful in determining
scaling exponents in experimental and 
numerical turbulent flow. In ESS, the p$^{th}$ order
longitudinal velocity structure
function
\be
  D_{p}^L (r) =
  \langle [
(\u (\x +\r) - \u (\x )) \cdot \e_r^L
  ]^p \rangle
\label{eq1}
\ee
is plotted against the third
order structure function.
Here, $\e_r^L$ is the unit vector in $\r$ direction; the unit vector
$\e_r^T$ used below is perpendicular to $\e_r^L$. 
The original motivation for picking the third
order structure function was the
Howard-v.\ Karman-Kolmogorov structure equation
\cite{my75,nel94,fri95}
\be
D_3^L (r) =- {4\over 5} 
\eps r + 6\nu {d\over dr} D_2^L (r),
\label{eq2}
\ee
saying that 
in the inertial subrange (ISR)
 $D_3^L(r)$ scales like $D_3^L \propto r$ and therefore
$D_p^L \propto r^{\zeta_p^L}\propto (D_3^L (r))^{\zeta_p^L }$
has the same scaling exponent as a function of $r$ or of $D_3^L$.
However, because of the poor statistical
convergence, rather than $D_3^L(r)$, the third order structure function
$D_3^{*L}(r)$,
calculated with the {\it modulus} of the velocity difference,
is
taken and it is argued
\cite{ben93b,bri94}
 that $D_3^{*L}(r)$ would also scale linearly with $r$ in
the ISR \cite{her95b}.
The resulting exponents,
which in general have to be distinguished
from the $\zeta_p^L$'s \cite{sto93b}, 
 are denoted as $\xi_p^L$, defined by
$D_p^L \propto (D_3^{*L})^{\xi_p^L}$, and they are found to be
remarkable universal, i.e., independent of
flow geometry and Reynolds number
\cite{arn96,bel96}.

Note that the degree of intermittency could be quantified by plotting
$D_p^L(r)$ vs {\it any} structure function $D_q^L(r)$,
(odd order moments taken
with the modulus, $D_q^{*L}(r)$).

In this paper we would like to demonstrate
that -- beyond the mere practical reason of better
statistics --
it seems really essential
for physical reasons
to take $D_3^{*L}(r)$ rather than
$D_3^L(r)$ 
to have ESS. We do so by examining ESS both for a full numerical
simulation \cite{gro97a} and for Batchelor's parametrization of the
structure function \cite{bat51}. This parametrization will give
 us the opportunity
to study finite Reynolds number $Re$ effects. We will
furthermore discuss 
the difference in scaling of the longitudinal as compared to
 the  transversal
structure functions.

We numerically solve the 3D incompressible Navier-Stokes equation
on a $N^3$ grid with periodic boundary conditions. A pseudospectral code is
used, the flow is forced on the largest length scales. For $N=96$ we achieve a
Taylor-Reynolds number $\rel = 110$; scales down to $3\eta$ are resolved;
$\eta = \nu^{3/4}/\eps^{1/4}$ is the Kolmogorov length, $\nu$ the kinematic
viscosity, $\eps$ the energy dissipation rate. Time integrations
 up to 150 large
eddy turnovers are performed; the flow is locally
isotropic to a high degree.
More details on the numerical flow
are given in ref.\ \cite{gro97a}.

%caption1
\begin{figure}[htb]
\setlength{\unitlength}{1.0cm}
\begin{picture}(9,9)
\put(0.5,0.5)
{\psfig{figure=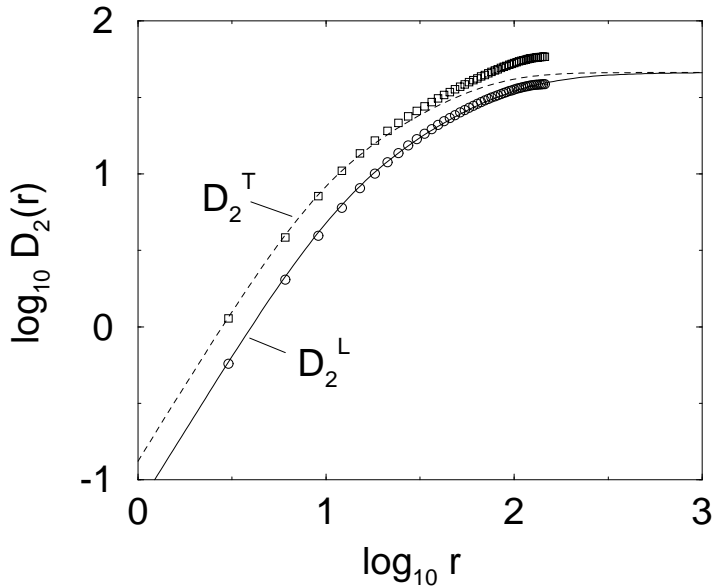,width=10cm,angle=-90}}
%\put(0.5,0.5){\psfig{figure=dummy.eps,width=6cm}}
\end{picture}
\caption[]{
Batchelor parametrization
(\ref{bat}) (solid line) of the N=96 data for the longitudinal
second order structure function $D_2^L(r)$ (numerical data:
circles). We chose $a=12.4\eta$
and $L=108\eta$.
Then we used eq.\ (\ref{eq6b}) to calculate $D_2^T(r)$ (dashed line)
which poorly compares with the numerical data (squares) for $r$
beyond the VSR.
}
\label{batchelor_data}
\end{figure}

\section{Batchelor's parametrization}
The longitudinal second order structure function $D_2^L(r)$ is shown in
figure \ref{batchelor_data}. As in the whole paper, lengths are given
in multiples of $\eta$ and velocities in multiples of
$(\eps \eta )^{1/3}$. An ISR
 scaling range is hardly developed because $\rel = 110 $ is
still small. The data are very well fitted by a parametrization of Batchelor's
type \cite{bat51,eff87,sto93,amg95,men96}
with an additional large scale cutoff $L$
\cite{amg95,amg96}
\be
D_2^L (r) = {\eps \over 15 \nu} {r^2 \over
\left[ 1+ \left( {r\over a}\right)^2\right]^{1-\zeta/2} }
{1\over \left[ 1+ \left(r\over L \right)^2\right]^{\zeta/2}} .
\label{bat}
\ee
Here, $\zeta=\zeta_2$ is the asymptotic ISR scaling exponents
which from our \cite{gro97a} and others' 
\cite{ben93b,ben94a,bri94,ben96b,fri95}  ESS analysis
we take to be $\zeta = 0.70$. 
The only
other free parameters are the
viscous subrange - inertial subrange (VSR-ISR) crossover scale $a$
and, of course, the large scale cutoff $L$. 
From a fit of eq.\ (\ref{bat})
to  the present numerical data we find $a=12.4\eta$  and
$L=108\eta$.

Note that we do not want to imply that all flow fields show a large
scale saturation of type (\ref{bat}). However, those data
(both numerical and experimental)
we analyzed
(see also  \cite{amg95,amg96})
were well described by eq.\ (\ref{bat}).
As eq.\ (\ref{bat}) guarantees an {\it analytic} behavior of both
the correction term to $r^2$ on the small scale side and
the correction term to
$r=const$ on the large scale side we think that this is not
accidental.

%caption1
\begin{figure}[htb]
\setlength{\unitlength}{1.0cm}
\begin{picture}(9,9)
\put(0.5,0.5)
{\psfig{figure=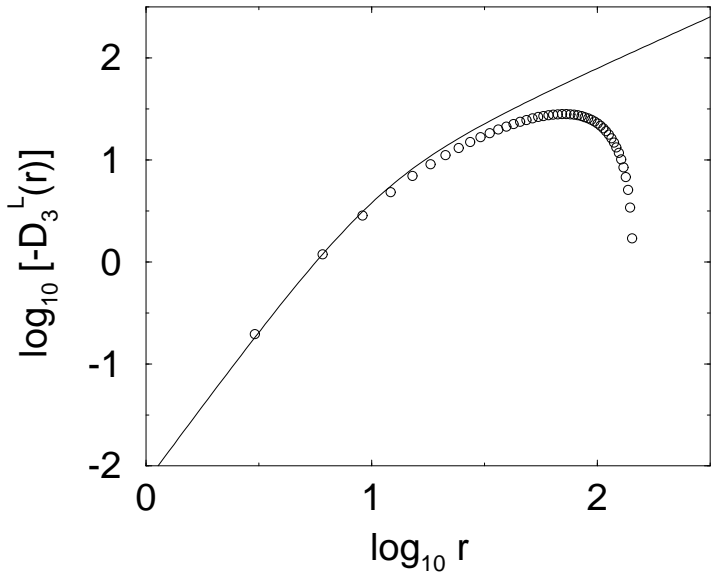,width=10cm,angle=-90}}
\end{picture}
\caption[]{
Third order structure function from our numerical simulation (circles),
compared with the one following from the Batchelor parametrization
of $D_2^L(r)$ (full line). The Taylor-Reynolds number is $\rel = 110$. 
}
\label{fig_d3}
\end{figure}

With the outer length scale $L$ we can define a Reynolds number
$Re = u_{1,rms} L/ \nu $.
For very large $Re$ the
structure function (\ref{bat}) develops an ISR scaling law
\be
D_2^L (r) = b^L (\eps r)^{2/3} \left( {r\over L}\right)^{\delta\zeta_2}. 
\label{eq5}
\ee
Here, $\dz_2 = \zeta_2 - 2/3$ is the scaling correction
to classical scaling 
and $b^L$ is often
called the
Kolmogorov constant. From our fit we have
$b^L = \eps^{1/3} L^{\dz_2} a ^{2-\zeta_2 } / (15\nu ) = 2.05$.
This value well agrees with the data $b^L = 1.6 - 2.5$
known from the literature
\cite{my75,eff87,nel94,fri95,sre95};
Sreenivasan \cite{sre95} gives $b^L =2.0 \pm
0.4$.
Note that the full structure function
$D_2(r) = 3D_2^L (r) + r {\partial D_2^L\over dr}$ (for isotropic flow)
asymptotically  scales with a law
of type (\ref{eq5}), too;
the prefactor is $b \approx 11 b^L /3 = 6$ -- $ 9$. Alternatively,
also $D_2 (r)$ can be fitted by a Batchelor parametrization
\cite{eff87,amg95} with similar quality \cite{amg95}. 

Our motivation to employ Batchelor's
 parametrization (\ref{bat}) is to be able
to {\it upscale} the second order structure function $D_2^L (r)$  to much
larger $Re$ (assuming that $b^L $ and $\zeta$ are fixed at $b^L = 2.0$ and
$\zeta = 0.70$) and thereby get consistent data for the
{\it transversal} second order structure function
\begin{equation}
  D_{p}^T (r) =
  \langle [
(\u (\x +\r) - \u (\x )) \cdot \e_r^T
  ]^p \rangle ,
\label{eq6}
\end{equation}
$p=2$, and for the third order
longitudinal structure function $D_3^L (r)$, which for
isotropic, homogeneous, incompressible
 turbulence both follow from $D_2^L (r)$, namely
through the relation
\be
D_2^T (r) = D_2^L(r) + {r\over 2} {d\over dr} D_2^L (r)
\label{eq6b}
\ee
and through eq.\ (\ref{eq2}), respectively.

To upscale $D_2^L (r)$ in eq.\ (\ref{bat}), we must know how the parameters $a$
and $L$ depend on the Reynolds number. If one accepts Sreenivasan's
observations that neither the (asymptotic) dimensionless energy dissipation
rate \cite{sre84,loh94a,gro95}
$c_\eps = {\eps L / u_{1,rms}^3}$,
nor the Kolmogorov constant $b^L$ \cite{sre95}
(but note also ref.\ \cite{pra94})
depend on the Reynolds number, one gets
a weak dependence of the VSR-ISR crossover on the Reynolds
number \cite{gro95},
\be
{a\over \eta} =  (15 b^L)^{3/(4-3\delta\zeta )}
\left( {\eta\over L} \right)^{3\dz/(4-3\dz)}.
\label{eq_a}
\ee
Once $b^L$, $\zeta$, and $Re$ are fixed,
$L$ and $\rel$ can easily be obtained from above
equations \cite{loh94a,gro95}.
With $D_2^L (\infty ) = 2u^2_{1,rms}$ we get
\be
{L\over \eta} = \left( { 2\over b^L }\right)^{3/8}  Re^{3/4},
\label{eq_l}
\ee
\be
\rel =
{\sqrt{15} u_{1,rms}^2 \over \sqrt{\nu \eps}}
=\sqrt{15}
\left({b^L\over 2} \right)^{3/4}  \sqrt{Re}. 
\label{eq_rel}
\ee
Vice versa, once $a$ and $L$ are known, we obtain $Re$ and $\rel$.
For above values the result is 
$Re = 520 $ and $\rel = 90$
for our numerical flow, in reasonable agreement to the direct numerical
result $\rel = 110$. As we will see,
the reason for the (modest) underestimation is that in the numerical flow
there are correlations left at the largest length scales.

\section{{$D_2^T $} and {$D_3^L  $} resulting from
Batchelor's parametrization}
For $D_2^T(r)$ we find poor agreement
between the curve evaluated from equation (\ref{eq6b}) and the 
numerically obtained values,
see figure \ref{batchelor_data}.
The reason is that at $r\sim L$ there
still is considerable correlation $\la \u(x+r) \u (x) \ra \ne 0$.
More precisely, at the maximal meaningful distance $r_{max}$ when
employing periodic boundary conditions, namely, when $r$ equals half of
the periodicity length (here, $r_{max}  \approx 146\eta$), we find
$\la u_j (\x+ \e_r^L r_{max}) u_j (\x )\ra / \la u_j^2 (\x ) \ra \approx 0.25$
and 
$\la u_j (\x+ \e_r^T r_{max}) u_j (\x )\ra / \la u_j^2 (\x ) \ra \approx -0.15$
for $j=1,2$, or $3$.
Therefore, $D_2^L (r_{max} )$ is smaller and 
$D_2^T (r_{max} )$ is larger than $2 \la u_j^2 \ra$,
 which is the value the
structure functions would take for perfect decorrelation between $\x$ and
$\x + \r_{max}$. Geometrically, the above correlations mean that there is an
eddy with diameter $r\sim r_{max}
\sim \pi L$ in the numerical flow.
The possibility of such large eddies
 is a consequence of the periodic boundary
conditions (in contrast to boundary conditions which put the velocity
to zero at the edge of the flow volume)
and will also survive for larger $\rel$. Such a large scale eddy
implies that the flow is {\it not} isotropic and homogeneous at the large scales
and therefore it should be no surprise that equation (\ref{eq6b}),
whose  derivation requires isotropy and homogeneity, does not lead to
good agreement with the data at large scales.
Note that the same problem occurs in experimental flow, see e.g.\
figure 1a of ref.\ \cite{her95}. At the largest measured distance
$r\approx 650 \eta$ (for $\rel = 300$) it clearly is $D_2^T (r) > D_2^L(r)$,
revealing a large scale eddy.

Also for 
$D_3^L(r)$
the agreement
between the
direct numerical values and those from the 
analytic equation (\ref{eq2}), which 
assumes isotropy and homogeneity,
is poor, see 
figure \ref{fig_d3}. The numerical
structure function $D_3^L(r)$ bends down for
large $r$ as the velocity differences at large scales have
Gaussian like statistics and
consequently odd order moments almost vanish. 
This feature which is not described by eq.\ (\ref{eq2}) is due to the boundary
effects; more precisely, because there are no larger eddies than on the 
scale $r_{max}$ which could provide correlations. 
In principle, this deficiency can be cured by adding a corresponding
term to that equation as e.g.\ done in ref.\ \cite{yak92}.

%caption1
\begin{figure}[htb]
\setlength{\unitlength}{1.0cm}
\begin{picture}(9,9)
\put(0.5,0.5)
{\psfig{figure=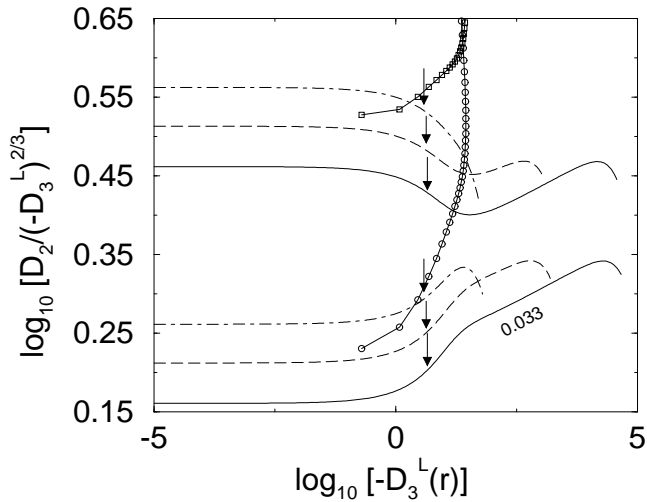,width=10cm,angle=-90}}
%{\psfig{figure=DISK_09:[detlef.rewa]ess_d2_bat.ps,width=10cm,angle=-90}}
%\put(0.5,0.5){\psfig{figure=dummy.eps,width=6cm}}
\end{picture}
\caption[]{
Compensated
ESS type plot  $D_2/(D_3^L)^{2/3}$
vs  $D_3^L$ for the structure functions
from the numerical simulation
(circles: longitudinal; squares: transversal)
and those 
following from the Batchelor parametrization (\ref{bat}) 
of $D_2^L$
for Reynolds numbers
$Re=5.2 \cdot 10^2$ as in the numerical simulation, for 
$Re=5.2 \cdot 10^4$, and for 
$Re=5.2 \cdot 10^6$, corresponding to Taylor-Reynolds numbers of
$\rel = 90$ (dot-dashed),
$\rel = 900$ (dashed),
and $\rel=9000$ (solid);
$\zeta=0.7$, $b^L = 2.0$. The three lower curves are for the
longitudinal structure functions, the three upper ones for the
transversal ones. 
The arrows indicate $10\eta$. The external length
scales $L$ are beyond the shown
regimes. 
}
\label{ess_d2_bat}
\end{figure}

%caption1
\begin{figure}[htb]
\setlength{\unitlength}{1.0cm}
\begin{picture}(9,9)
\put(0.5,0.5)
{\psfig{figure=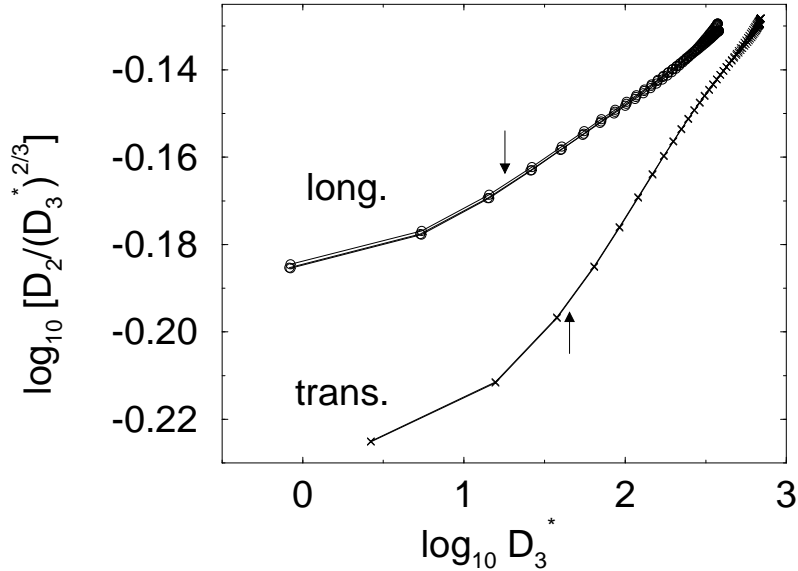,width=10cm,angle=-90}}
%\put(0.5,0.5){\psfig{figure=dummy.eps,width=6cm}}
\end{picture}
\caption[]{
Compensated
ESS type plots
$D_2^L/(D_3^{*L})^{2/3}$ vs  $D_3^{*L}$ (three different data sets
for three different space directions which, however, agree very well)
and $D_2^T/(D_3^{*T})^{2/3}$ vs  $D_3^{*T}$ (two data sets, also agreeing)
for the numerical turbulence, $\rel = 110$.
The arrows again indicate $10\eta$; the data points are for
$r=3\eta$,
$r=6\eta$,
$r=9\eta , \dots$, left to right.
}
\label{ess_d2_stern}
\end{figure}

\section{How to apply ESS?}
We now plot the second vs the third order structure function in a
compensated ESS type plot,
i.e., $D_2^{L} / (D_3^L)^{2/3}$ vs $D_3^L$, see figure \ref{ess_d2_bat}. 
The scaling regime in the numerical simulation
is by far too short to identify any scaling exponent.
However, if we repeat this plot, but now with $D_3^{*L}$ rather than
$D_3^L$, ESS {\it is} seen, see figure \ref{ess_d2_stern}. The reason is
that $D_2^L$ and $D_3^{*L}$ have the same type of large scale saturation
(i.e., becoming constant), whereas $D_3^L$ has a different type of large
scale behavior (namely, dropping to zero).
We have to conclude that 
the extension
of the scaling regime by using
ESS is mainly an extension towards {\it large} scales.
This even holds if $\rel$ is so small that there is no ISR yet. 
The {\it small scale onset}
 of scaling still is around $r\sim 10\eta$, whether
plotting $D_2$ vs $D_3^{*L}$ or vs $r$, which is roughly the crossover
scale $a\approx 12.4\eta $ found from employing eq.\ (\ref{bat}).
In their numerical simulation Briscolini et al.\ \cite{bri94}
find an ESS extension down to $r \sim 7\eta$ (see figure 4 of their paper),
roughly the same as the $10\eta$ reported here, but slightly smaller.
As pointed out to us by R.\ Benzi, the origin of the slight difference
may be that the small scale resolution in Brisolini et al.\ \cite{bri94}
is down to $1\eta$, whereas here we only have a $3\eta$ resolution
and the deviations in the structure functions due to the lower end of the
resolution may influence the
scaling exponents in a certain range of larger scales.  

\begin{figure}[htb]
%\begin{figure}[p]
\setlength{\unitlength}{1.0cm}
\begin{picture}(9,9)
\put(-0.,7.7){\large{a)}}
\put(-0.,4.){\large{b)}}
\put(0.5,0.5)
{\psfig{figure=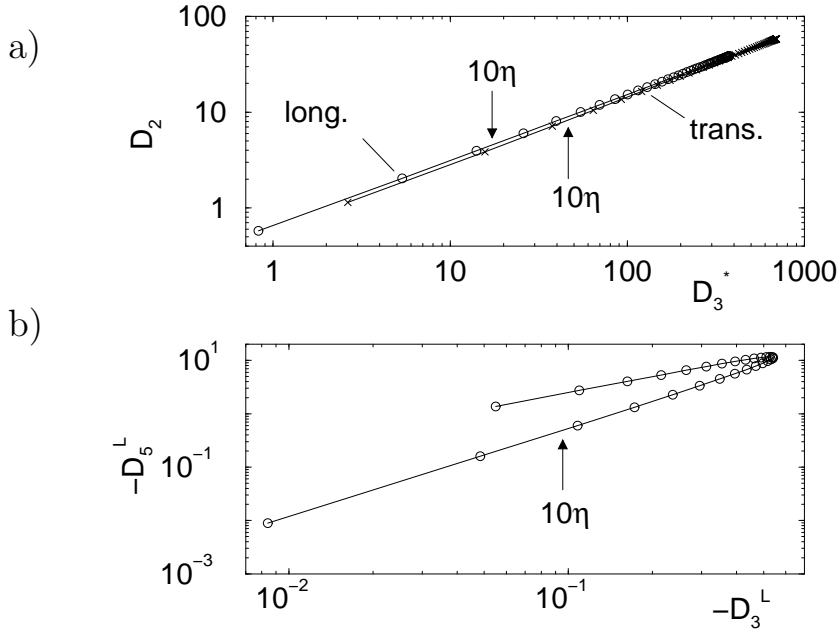,width=10cm,angle=-90}}
%\put(0.5,0.5){\psfig{figure=dummy.eps,width=12cm}}
\end{picture}
\caption[]{
(a) ESS plots $D_2(r)$ vs $D_3^*(r)$ for the longitudinal and
transversal structure functions; $\rel= 110$. The data are the same as in
the previous figure, where we showed the {\it compensated} ESS type plots,
in which the different behavior in the VSR and the ISR is clearly visible.
\\
(b) ESS plots for $-D_5^L(r)$ vs $-D_3^L (r)$.
In this figure the Taylor-Reynolds number is only $70$, but we
checked
that the lack of ESS does not decrease with increasing $\rel$. 
}
\label{fig_odd}
\end{figure}

Our next point is to
advocate {\it compensated} ESS plots for the visualization
of intermittency effects. Already 
Meneveau \cite{men96} -- see figure 1 of that paper -- reveals
how misleading an ESS plot $D_p^L$ vs $D_3^{*L}$
can be.
Here, we demonstrate this in figure \ref{fig_odd}a which shows the
original ESS plot $D_2$ vs $D_3^*$, only {\it pretending} better scaling
than in figure \ref{ess_d2_stern}. 
The reason is that in the VSR the scaling exponent is $2/3$ for
trivial reasons, a value which can hardly be distinguished by eye
from the ISR value $2/3 + \delta\zeta_2 \approx 0.70$. 
To avoid this similarity of the VSR and ISR exponents,
 we prefer to use {compensated} ESS plots
\cite{gro96,gro97a}.

%caption1

\begin{figure}[htb]
%\begin{figure}[p]
\setlength{\unitlength}{1.0cm}
\begin{picture}(9,10)
\put(-0.,7.7){\large{a)}}
\put(-0.,4.){\large{b)}}
\put(0.5,0.5)
{\psfig{figure=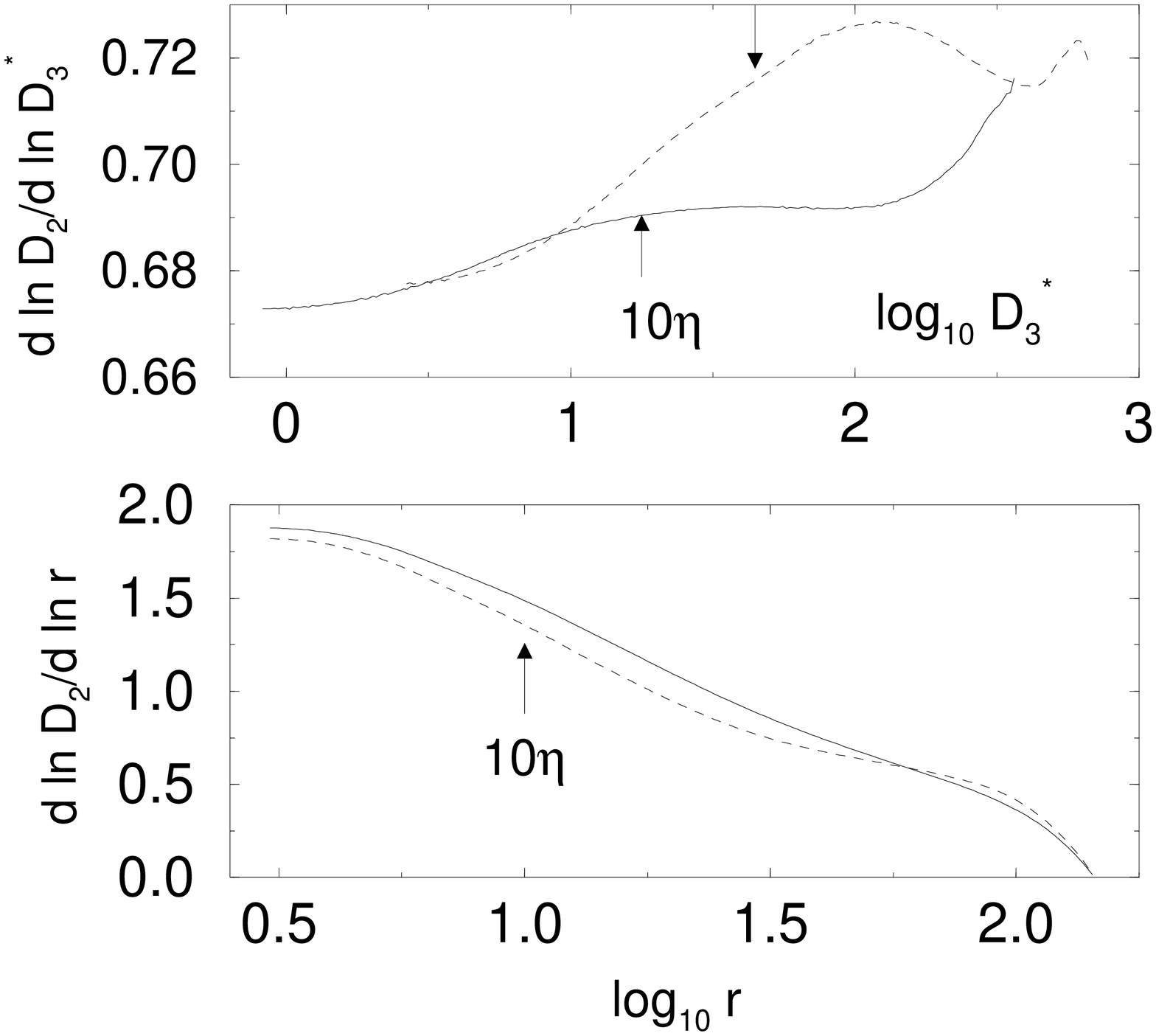,width=10cm,angle=-90}}
%\put(0.5,0.5){\psfig{figure=dummy.eps,width=12cm}}
\end{picture}
\caption[]{
(a) Local slopes
$d\lg D_2^{L}/d\lg D_3^{*L}$ (solid) and 
$d\lg D_2^{T}/d\lg D_3^{*T}$ (dashed)
of the curves in figure \ref{fig_odd}a. The arrows indicate $10\eta$.
\\
(b) Local slopes
$d\lg D_2^{L}/d\lg r$ (solid) and 
$d\lg D_2^{T}/d\lg r$ (dashed)
of the numerical data
curves $D_2^{L,T}(r)$
in figure \ref{batchelor_data}.
}
\label{fig_local_slopes}
\end{figure}

If one plots {\it local slopes}
\cite{sto93b} as done in figure \ref{fig_local_slopes}a
it of course does not make any difference whether one takes them
from compensated ESS plots or standard ESS plots. From figure
\ref{fig_local_slopes}a one notices that the ESS scaling
$\xi_2^L \approx 0.69$ and $\xi_2^T \approx 0.72$ 
begins around
$10\eta$. From figure \ref{fig_local_slopes}b one also notices that
without ESS one could not deduce any scaling exponent at all for the
small Reynolds number of our numerical calculation.
ESS is thus useful already for the simple reason that a transition
from a local slope of 2/3 to roughly 0.70 is {\it shorter} than
from a local slope 2 to roughly 0.70.

%caption1

As we will show now, there is no 
extended scaling regime towards scales much smaller than order $\sim 10\eta$,
either, if one does ESS type plots with $D_3^L$ instead of $D_3^{*L}$.
We do so by plotting
$D_2^L / (D_3^L)^{2/3}$ vs $D_3^L$ with $D_3^L$ following (via eq.\
(\ref{eq2}))
from
the Batchelor parametrization eq.\ (\ref{bat}) of $D_2^L$ for various $Re$,
see figure \ref{ess_d2_bat}. 
We observe three regimes: The VSR without any scaling corrections (i.e.,
a horizontal line in figure \ref{ess_d2_bat}),
a crossover regime, corresponding to the range from $r\sim 1\eta$ to
$r \sim 10\eta$, and only for large scales  and large $Re \ageq 500$
the ISR scaling
corrections $\dz_2 = 0.033$ can be identified.

To summarize this subsection: There seems to be ESS towards large
scales, if the structure functions plotted against each other are both
calculated with the moduli, i.e., have the same large scale saturation
behavior.
In particular, for the third order longitudinal
structure function this means that
it is essential to take $D_3^{*L}$ rather than $D_3^L$ and to clearly
distinguish between the $\zeta_p$ and $\xi_p$ exponents.
This was already stressed by Stolovitzky and Sreenivasan \cite{sto93b},
see in particular their figures 2 and 4 where they compare local slopes
of $D_8^L$ vs.\ $D_3^L$ and $D_3^{*L}$.

The natural question to ask is: Is there also ESS for odd order
structure functions
(calculated {\it without} the modulus)
plotted against each other?
In figure \ref{fig_odd}b we plot $-D_5^L$ vs $-D_3^L$. No ESS towards
large scales is seen. It seems that odd and even order moments
obey fundamentally different types of statistics. This finding may
be connected to Herweijer and van der Water's finding \cite{her95c}
that $\zeta_p$ for {\it odd} $p$ (calculated from $D_p^L (r) \propto
r^{\zeta_p}$) are {\it smaller} than expected from an extrapolation
of the neighboring $\zeta_{p\pm 1}$ (for which the $p\pm 1$ are even). 
A difference between the $\zeta_p$ for odd and even $p$ was also
found by Stolovitzky et al.\ \cite{sto93,sto93b}; however, for the flow
analyzed in that references the $\zeta_p$ for odd $p$ are {\it larger}
than expected from the extrapolation.

We do not understand {\it why} nonuniversal forcing and large scale
boundary effects roughly cancel out in even order structure functions
but not in odd order ones (calculated without the modulus). With a more
elaborate technique it may even be possible to extract nonuniversal
properties also from ESS plots of even structure functions. 
On the other hand, we cannot exclude that there is more universality
in {\it decaying} turbulence where less anisotropy through the forcing and the
boundaries is felt.

\section{Longitudinal vs transversal structure functions}
Next, we focus on the difference in the scaling between
longitudinal and transversal structure functions,
$D_p^L(r) \propto r^{\zeta_p^L}$ and 
$D_p^T(r) \propto r^{\zeta_p^T}$, respectively. Recently, different degrees
of intermittency
for longitudinal and transversal fluctuations
were reported 
in some experiments \cite{her95,her97,cam97} and numerical simulations
on decaying turbulence \cite{bor97}. We confirmed these findings 
for statistically stationary turbulence \cite{gro97a} (see also
ref.\ \cite{chen}).
More precisely, it were
the ESS type scaling exponents $\xi_p^L$ and $\xi_p^T$, defined by
$D_p^L \propto (D_3^{*L})^{\xi_p^L}$ and 
$D_p^T \propto (D_3^{*T})^{\xi_p^T}$, which are clearly different;
we found $\dx_6^L = 0.21 \pm 0.01$ and $\dx_6^T = 0.43\pm 0.01$ for the
deviations from the mean field value $\xi_6 = 2$ \cite{gro97a}.

%caption1
%\begin{figure}[htb]
\begin{figure}[p]
\setlength{\unitlength}{1.0cm}
\begin{picture}(9,17)
\put(-0.,7.){\large{b)}}
\put(0.5,0.5)
{\psfig{figure=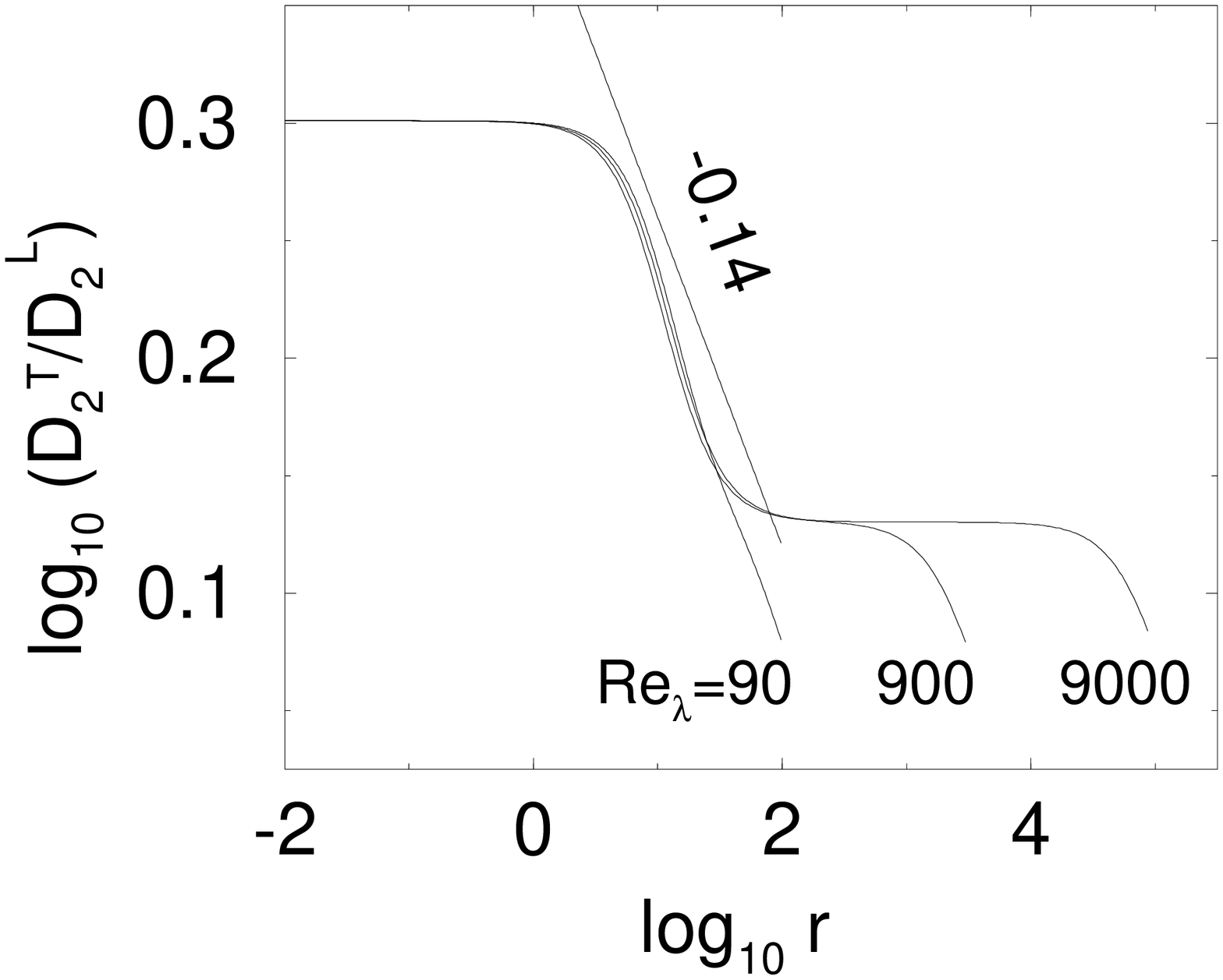,width=11.5cm,angle=-90}}
\put(-0.,16.5){\large{a)}}
\put(0.5,9.5)
{\psfig{figure=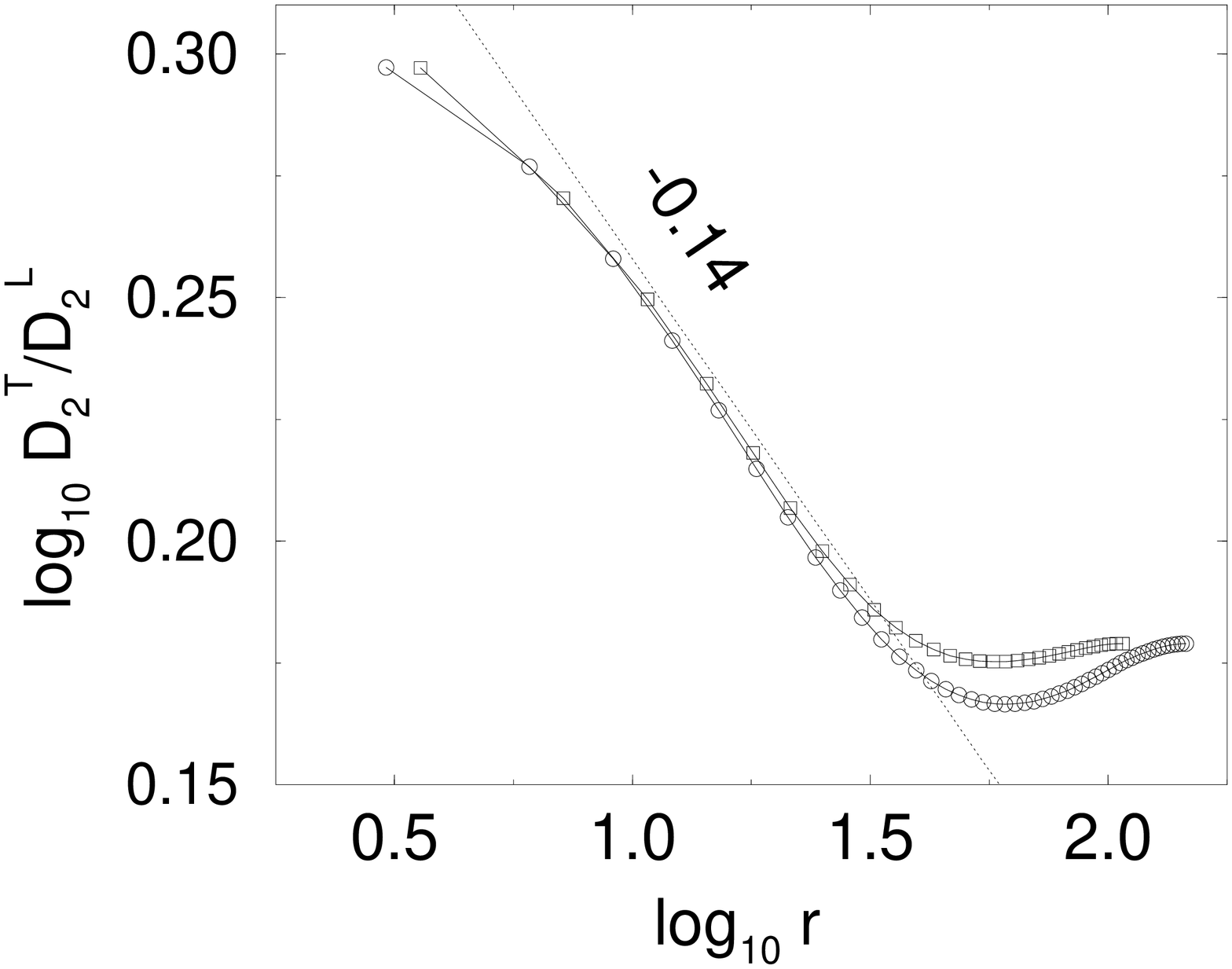,width=10cm,angle=-90}}
\end{picture}
\caption[]{
(a) The ratio 
$D_2^T(r)/D_2^L(r)$ vs  $r$ for the numerical data of  a
$\rel =110$ (circles) 
and a $\rel=70$ (squares)
simulation. {\it Erroneously}, one may deduce {\it different}
scaling of $D_2^L(r)$ and $D_2^T(r)$ because the ratio depends on $r$.
We also included an apparent slope of $-0.14$.\\
(b)
$D_2^T(r)/D_2^L(r)$ as a function of $r$ for the Batchelor parametrization
(\ref{bat})
for Taylor-Reynolds numbers of
$\rel = 90$,
$\rel = 900$,
and $\rel=9000$. The plateau in the ISR (at
roughly $lg(4/3) = 0.125$)
expected for isotropic homogeneous flow
only starts to develop for as large $\rel$ as $\rel \sim 500$.
The VSR plateau is at $lg ~2 = 0.30$.
The $r$-dependent intermediate range characterizes the transition
between  the VSR and the ISR. The slope is related to its width
and the different heights of the two plateaus.
As a side remark we mention that in this plot one can also notice
the $\rel$ dependence of the VSR-ISR crossover $a$, cf.\ eq.\ (\ref{eq_a}):
the transition range
is shifted towards smaller $r$ with increasing $\rel$.
}
\label{ess_tdl_bat}
\end{figure}

One would be tempted to conclude that the best way to see a deviation
in scaling between $D_p^L$ and $D_p^T$ would be to plot the {\it ratio}
$D_p^T / D_p^L$ vs $ r$  (or vs $D_3^{L}$).
According to eq.\ (\ref{eq6b}), $D_2^T (r)$ and $D_2^L (r)$ scale
the same in the ISR, i.e., the ratio should be constant. However,
figure \ref{ess_tdl_bat}a
seems to imply different scaling of $D_2^L(r)$ and $D_2^T(r)$.

The reason for this apparent discrepancy is that 
the argument of equal scaling of $D_2^L(r)$ and $D_2^T(r)$ is only
valid if both structure functions scale individually. This is  not
the case in the transition ranges or if there is not any ISR yet. Here,
the Reynolds number achieved in the full simulation is by far
too small to give the asymptotic (ISR)
scaling. In figure \ref{ess_tdl_bat}b we redo
this type of plot, but now within the Batchelor parametrization for which we can
achieve arbitrarily large $\rel$. 
Only if $\rel \ageq 500$ a plateau starts to develop,
showing the onset of the asymptotically correct
ISR behavior.
To reliably determine scaling exponents from the plateau, one would
need at least Taylor-Reynolds numbers $\sim 1000$ and beyond. 
For $\rel \sim 100$
there is a fake scaling law with an apparent exponent of $-0.14$,
which has nothing to do with inertial range scaling.

Going back to eqs.\ (\ref{bat}) and (\ref{eq6b}), this behavior can be
understood.
In the VSR we must have
$D_2^T / D_2^L   = 2$
(because of eq.\ (\ref{eq6b}) and $D_2^L \propto r^2$)
and in the ISR we have 
$D_2^T / D_2^L   \approx 4/3$
(because of eq.\ (\ref{eq6b}) and roughly $D_2^L \propto r^{2/3}$),
just as seen in figure \ref{ess_tdl_bat}b.
The crossover between these two regimes is
about a decade. The same can be seen
from figure \ref{ess_d2_bat} where besides
$D_2^L / (D_3^L)^{2/3}$ we also plotted
$D_2^T / (D_3^T)^{2/3}$
vs $D_3^L$. In the crossover regime where the former curve bends up, the latter
bends down. Again, only for $\rel \ageq 500$ the asymptotic scaling exponent
$\dz_2^L = \dz_2^T = 0.033$ starts to be observable.

The same finite $\rel$ effects which we discussed for the 2nd order structure
functions,
 where $D_2^L$ and $D_2^T$ are {\it known} to have the same scaling, 
will {\it hinder} to determine  scaling exponents vs $r$ (or vs
$D_3^L$)
in higher order structure functions for too low $\rel \aleq 500$.

\section{Summary} 
To conclude,
we confirmed the finding of Briscolini et al.\ \cite{bri94} that
ESS does not extend to scales below order $\sim 10 \eta$. 
We furthermore showed from
calculations with the Batchelor parametrization
that scaling exponents
$\zeta_p$ calculated from structure functions plotted vs $r$ (or 
vs $D_3^L$) can only securely be measured for
$\rel$ sufficiently larger than
$ 500$. For smaller $\rel$, in particular for all present day
numerical
simulations, one is restricted to {\it relative}, ESS type scaling exponents
$\xi_p$ calculated from ESS type plots 
$D_p^{*L}(r)$ vs
$D_q^{*L}(r)$ and 
$D_p^{*T}(r)$ vs
$D_q^{*T}(r)$,
whereby it is essential to calculate the structure functions from
the moduli of the velocity differences.
For odd order moments, calculated {\it without} taking the modulus,
ESS does not hold in the presented numerical simulation.

\vspace{1cm}

\noindent
{\bf Acknowledgements:}
We thank Roberto Benzi and
Luca Biferale for very helpful comments on the manuscript.
Support for this work by
the Deutsche Forschungsgemeinschaft (DFG) under grant SBF185 and by
the German-Israel Foundation (GIF) is gratefully acknowledged.
The HLRZ J\"ulich supplied us with computer time. 

\vspace{0.5cm}

\noindent
 e-mail addresses:\\
grossmann$\_$s@physik.uni-marburg.de\\ 
lohse@stat.physik.uni-marburg.de\\
reeh@mailer.uni-marburg.de

%\bibliography{literatur}

\begin{thebibliography}{10}

\bibitem{ben93b}
R. Benzi {\it et~al.},
Phys. Rev. E {\bf 48},  R29  (1993).

\bibitem{ben94a}
R. Benzi, S. Ciliberto, C. Baudet, and G.~R. Chavarria, Physica D {\bf 80},
  385  (1995).

\bibitem{bri94}
M. Briscolini, P. Santangelo, S. Succi and R. Benzi,
Phys. Rev. E {\bf 50},  R1745  (1994).


\bibitem{my75}
A.~S. Monin and A.~M. Yaglom, {\em Statistical Fluid Mechanics} (The MIT Press,
  Cambridge, Massachusetts, 1975).

\bibitem{nel94}
M. Nelkin, Advances in Physics {\bf 43},  143  (1994).

\bibitem{fri95}
U. Frisch, {\em Turbulence} (Cambridge University Press, Cambridge, 1995).

\bibitem{her95b}
Roberto Benzi (priv.\ communication) communicated us that for an experimental
flow in a regime where no forcing is felt this indeed were the case.
Other authors find differences; e.g., J. A.
Herweijer
(Ph.D. thesis, University of Eindhoven, 1995)
finds $D_3^L (r) \propto r^{1.035 \pm 0.005}$ and
$D_3^{*L} (r) \propto r^{1.055 \pm 0.005}$ for a wind tunnel measurement.


\bibitem{sto93b}
G. Stolovitzky and K. R. Sreenivasan,
Phys. Rev. E 48, R33 (1993).

\bibitem{arn96}
A.~Arneodo et~al., Europhys. Lett. {\bf 34},  411  (1996).

\bibitem{bel96}
F. Belin, P. Tabeling, and H. Willaime, Physica D {\bf 93},  52  (1996).

\bibitem{gro97a}
S. Grossmann, D. Lohse, and A. Reeh,
``Different intermittency for longitudinal and transversal 
turbulent fluctuations'', submitted to Phys. Fluids 
(chao-dyn/9704014).

\bibitem{bat51}
G.~K. Batchelor, Proc. Camb. Philos. Soc. {\bf 47},  359  (1951).

\bibitem{eff87}
H. Effinger and S. Grossmann, Z. Phys. B {\bf 66},  289  (1987).

\bibitem{sto93}
G. Stolovitzky, K.~R. Sreenivasan, and A. Juneja, Phys. Rev. E {\bf 48},  R3217
   (1993);
L. Sirovich, L. Smith, and V. Yakhot, Phys. Rev. Lett. {\bf 72},  344  (1994).

\bibitem{amg95}
D. Lohse and A. M\"uller-Groeling, Phys. Rev. Lett. {\bf 74},  1747  (1995).

\bibitem{men96}
Ch. Meneveau, Phys. Rev. E {\bf 54},  3657  (1996).

\bibitem{amg96}
D. Lohse and A. M\"uller-Groeling, Phys. Rev. E {\bf 54},  395  (1996).

\bibitem{ben96b}
R. Benzi {\it et~al.}, Physica D {\bf 96},  162  (1996).

\bibitem{sre95}
K.~R. Sreenivasan, Phys. Fluids {\bf 7},  2778  (1995).

\bibitem{sre84}
K.~R. Sreenivasan, Phys. Fluids {\bf 27},  1048  (1984).

\bibitem{loh94a}
D. Lohse, Phys. Rev. Lett. {\bf 73},  3223  (1994).

\bibitem{gro95}
S. Grossmann,
Phys. Rev. E {\bf 51},  6275  (1995);
G. Stolovitzky and K. R. Sreenivasan,
Phys. Rev. E {\bf 52},  3242  (1995).


\bibitem{pra94}
A. Praskovsky and S. Oncley, Phys. Fluids A {\bf 6},  2886  (1994).

\bibitem{her95}
J. Herweijer and W. van~der Water,  in {\em Advance in Turbulence V}, edited
  by R. Benzi (Kluwer Academic Publishers, New York, 1995), p.\ 210.

\bibitem{yak92}
V. Yakhot, 
Phys. Rev. Lett. {\bf 69},  769  (1992).

\bibitem{gro96}
S. Grossmann, D. Lohse, and A. Reeh,
Phys. Rev. Lett. {\bf 77},  5369  (1996).

\bibitem{her95c}
J. Herweijer and W. van~de Water, Phys. Rev. Lett. {\bf 74},  4651  (1995).

\bibitem{her97}
W. van~de Water and 
J. Herweijer, 
``High order structure functions of turbulence'', preprint, Eindhoven 1996;
W. van de Water and
J. A. Herweijer, Phys. Scripta T67, 136 (1996).

\bibitem{cam97}
R. Camussi and R. Benzi, Phys. Fluids {\bf 9},  257  (1997).

\bibitem{bor97}
O.~N. Boratav and R.~B. Pelz, Phys. Fluids {\bf 9},  1400  (1997).

\bibitem{chen}
S. Chen, K. R. Sreenivasan, M. Nelkin, and N. Cao,
``A refined similarity hypothesis for transversal structure functions'',
preprint (1997).

\end{thebibliography}

\end{document}